\documentclass[epjCONF]{svjour}
\usepackage{graphics}
\usepackage[varg]{txfonts} 
\usepackage[latin1]{inputenc}
\session-title{FUSION11}
\begin{document}
\title{$^{12}$C+$^{16}$O sub-barrier radiative capture cross-section measurements}
\author{A. Goasduff\inst{1,2} \fnmsep\thanks{\email{alain.goasduff@iphc.cnrs.fr}} \and S. Courtin
\inst{1,2} \and F. Haas \inst{1,2} \and D. Lebhertz \inst{3} \and D.G. Jenkins \inst{4} \and C. Beck \inst{1,2} 
\and J. Fallis \inst{5} \and C. Ruiz \inst{5}\and D. A. Hutcheon \inst{5}\and P.-A. Amandruz \inst{5}\and C. Davis \inst{5}\and U. Hager \inst{5}\and D. Ottewell \inst{5}\and G. Ruprecht \inst{5} }
\institute{Universit\'e de Strasbourg, IPHC, 23 rue du Loess  67037 Strasbourg, France
\and CNRS, UMR7178, 67037 Strasbourg, France
\and GANIL, CEA/DSM-CNRS/IN2P3, Bd Henri Becquerel, BP 55027, F-14076 Caen Cedex 5, France
\and Department of Physics, University of York, Heslington, York YO10 5DD, United Kingdom 
\and TRIUMF, 4004 Wesbrook Mall, Vancouver, British Columbia, V6T 2A3 Canada}
\abstract{
We have performed a heavy ion radiative capture reaction between two light heavy ions, $^{12}$C and $^{16}$O, leading to $^{28}$Si. The present experiment has been performed below Coulomb barrier energies in order to reduce the phase space and to try to shed light on structural effects. Obtained $\gamma$-spectra display a previously unobserved strong feeding of intermediate states around 11 MeV at these energies. This new decay branch is not fully reproduced by statistical nor semi-statistical decay scenarii and may imply structural effects. Radiative capture cross-sections are extracted from the data.
} 
\maketitle
\section{Introduction}

The isotope $^{28}$Si can be considered as a key nucleus to understand the coexistence between mean field effects and cluster structures. Indeed different structures coexist in its excited states even at low excitation energies. Recent antisymmetrized molecular dynamics (AMD) calculations for $^{28}$Si \cite{tani} have shown that cluster structures such as $^{12}$C-$^{16}$O or $^{24}$Mg-$\alpha$ may have large contributions for normal-deformed and superdeformed states, respectively. By using the heavy-ion radiative capture (HIRC) mechanism in which the compound nucleus decays solely by $\gamma$-ray emission we have investigated the $^{12}$C-$^{16}$O cluster structure in $^{28}$Si. Moreover this mechanism will directly populate $^{28}$Si at energies where the calculated nuclear matter densities have similar asymmetries as the $^{12}$C+$^{16}$O reaction \cite{ichi}.

The excitation function of $^{12}$C+$^{16}$O exhibits narrow resonances \cite{sand} around the Coulomb Barrier ($V_B\sim 7.8$ MeV). These resonances are correlated in all reaction channels including the HIRC. Baye and Descouvemont \cite{baye} have shown, using GCM calculations, that for the similar system $^{12}$C+ $^{12}$C, several observed resonances can be interpreted in terms of molecular resonances. Although HIRC is a rare process compared to the dominant fusion-evaporation channels (n, p, $\alpha$), $\gamma$-decay will favourably populate states with large structural overlaps with the entrance-channel thus allowing us to try to identify the possible cluster states. 

\section{Experiment}

Our experimental program is focused on the study of HIRC in the $^{12}$C+$^{12}$C and $^{12}$C+$^{16}$O \cite{doro,jenk} systems. The present $^{12}$C($^{16}$O,$\gamma$) reaction has been performed at the TRIUMF facility (Vancouver, Canada). A $^{16}$O ISAC beam has been used at two resonant energies down to 15\% below $V_B$ ($E_{lab}$ = 1.07 and 0.96 AMeV) on high purity (99.9\%) $^{12}$C thin ($50\ \mu g.cm^{-2}$) targets. The DRAGON recoil separator has been used at 0$^\circ$ to select the $^{28}$Si recoils. We have used the BGO array to record the $\gamma$-rays in coincidence with the heavy recoils identified in a DSSSD placed at the DRAGON focal plane.

\section{Results and discussion}
\subsection{$\gamma$-ray decay}

\begin{figure}[h]
\resizebox{\columnwidth}{!}{\includegraphics{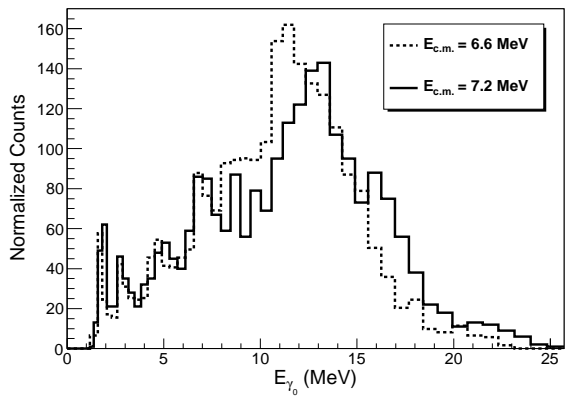} }
\caption{Highest energy $\gamma$-ray  spectra in coincidence with $^{28}$Si for the $^{12}$C+$^{16}$O reaction at $E_{c.m.}$= 6.6 MeV (dashed line) and $E_{c.m.}$= 7.2 MeV (full line). Spectra are normalized to the same integral.}
\label{spectra_e0}
\end{figure}

Fig.~\ref{spectra_e0} displays the highest energy $\gamma$-ray in each $\gamma$-event recorded in the BGO array in coincidence with the $^{28}$Si for the two studied energies. Both spectra show the same global structure. Concerning the region above 17 MeV, the direct transition to the ground state (g.s.) contributes to the bump above 20 MeV but due to the response function of the spectrometer and the deviation out of 0$^{\circ}$ of the recoils induced by high energy $\gamma$-ray emission very few recoils enter DRA\-GON. The feeding of the $2^+_1$ (1.778 MeV) from the resonances is also observed ($E_{\gamma_{0}} \sim 21$ MeV) at both energies. For the lowest energy we also have a $\gamma$-line around 18 MeV corresponding to the feeding of the $4^+_1$ (4.617 MeV).

 Concerning the low energy part of the spectra, the decay of the two first excited states of the $^{28}$Si can be identified ($E_{\gamma_{0}}$ = 1.78 and 2.84 MeV). The decay of the prolate head band $0^+_3$ (6.690 MeV) to the $2^+_1$ contributes to the line around 5 MeV but this line does not correspond to a single state decay. We observe also the decay of the $3^-_1$ to the g.s. which contributes to the $\gamma$-line at 6.8 MeV, a direct decay from the resonance is also observed for the two energies. It has been shown at higher energies \cite{doro} that this first state of negative parity in $^{28}$Si is crucial to understand the $\gamma$-spectra.
 
Due to the low resolution of the BGO $\gamma$-array and the large number of states in $^{28}$Si around 10 MeV, the large bump between 10 and 15 MeV with correspond to the feeding of intermediate states around 8-13 MeV cannot be discussed in terms of the feeding of particular states. Moreover in this region the transitions from the resonances and the subsequent decays to the g.s. are quite close in energy. This previously unobserved $\gamma$-flux at these reaction energies has to be linked to what has been observed at higher energies ($E_{c.m}$ = 8.5, 8.8 and 9.0 MeV) \cite{doro}, where the bump was around 14 MeV. Taking into account that in our previous experimental campaign the $^{28}$Si was populated at higher excitation energies ($E^* \sim 25.3$, 25.6 and 25.8 MeV), than in the present experiment ($E^* \sim 23.3$ and 23.9 MeV), the observed differences on the bump centroids are only resulting from the differences in entrance-channel energies.  

\begin{figure}[h]
\resizebox{\columnwidth}{!}{\includegraphics{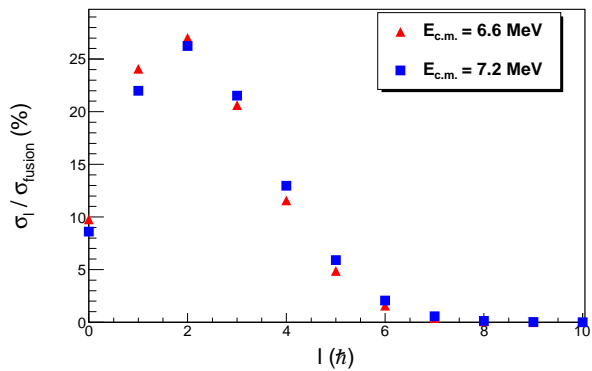} }
\caption{Spin distributions normalized to fusion cross-sections obtained by CCFULL \cite{ccfull} calculations for the two studied energies: red triangles correspond to $E_{c.m.}$ = 6.6 MeV and blue squares to $E_{c.m.}$ = 7.2 MeV. }
\label{spin_distri}
\end{figure}

\begin{figure}[h]
\resizebox{\columnwidth}{!}{\includegraphics{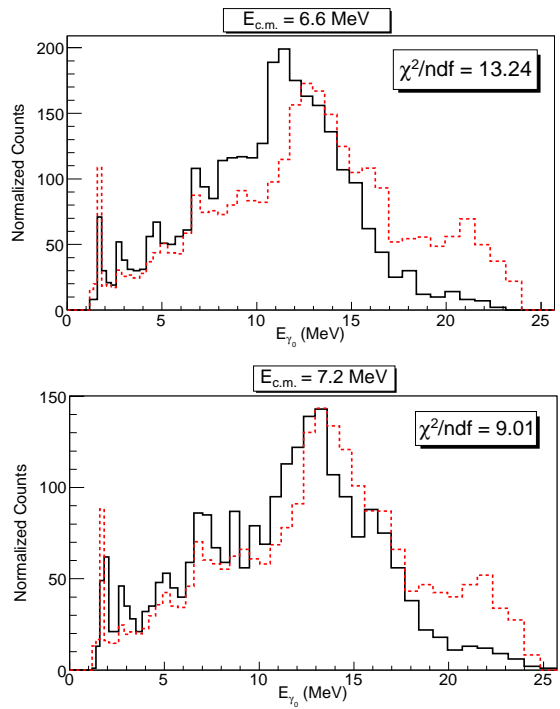} }
\caption{Highest energy $\gamma$-ray spectra in coincidence with $^{28}$Si for $E_{c.m.}$= 6.6 MeV (full line) $E_{c.m.}$= 7.2 MeV (full line) compared to fully statistical decay (red dashed line) using the spin distributions given in Fig.~\ref{spin_distri}. Simulation spectra are normalized to the data integrals.}
\label{simu_spin_distri}
\end{figure}

In order to fully understand the decay pattern of the resonances and see if we have a structural effect which will favourably feed particular states, we compared our $\gamma$-spectra to GEANT simulations with different conditions in the entrance-channel. All simulations include 68 known bound or quasi bound states of $^{28}$Si between 0 and 13 MeV and their $\gamma$-decays. Above the particle threshold ($\sim$10 MeV for the lowest one, in this case $\alpha$+$^{24}$Mg) we have selected states with a large $\Gamma_\gamma$ compared to particle emission. To estimate the branching ratios of the entrance-channel to each of these states we have used Weisskopf estimates and the reported average strengths of electric and magnetic transitions in $^{28}$Si \cite{endt}. In this self-conjugate nucleus ($T_z = 0$), some particular selection rules on isospin apply for L=1, $\Delta T$=0 transitions: E1 are forbidden and M1 strengths are reduced by a factor of 100. As we use  an entrance-channel with $T=0$ in all simulations, these rules will strongly influence the obtained $\gamma$-spectra.

We have first simulated a fully statistical scenario, which consists of a spin distribution in the entrance-channel. Spin distributions for the two studied energies are given in Fig.~\ref{spin_distri}. These distributions are the results of coupled-channel calculations, performed with the CCFULL code \cite{ccfull}, and are obtained by adjusting the diffuseness parameter ($a=$0.57 and 0.55 for $E_{c.m.}$=6.6 and 7.2 MeV) in order to reproduce the fusion cross-sections from \cite{cujec} and given in Fig.~\ref{plot_xsection}. The two curves are normalized to the calculated fusion cross-sections. Below $V_B$, phase space is reduced and spin distributions are quite narrow and centered at low spin ($2^+$) compared to what was obtained at higher energies for the $^{12}$C+$^{16}$O reaction \cite{doro}.

Simulated spectra (dashed lines) are given in Fig.~\ref{simu_spin_distri} along with experimental data (full lines). For the two energies we globally reproduce the $\gamma$-spectrum shape. Indeed a large bump between 11 and 15 MeV dominates the spectrum. However looking into the details we see that this statistical scenario is unable to reproduce the intensities of the different transitions between the low-lying states. Another discrepancy concerns the high energy part of the simulations where the direct feeding of the two first excited states is overestimated. Concerning the intermediate region of the spectra the bump is better reproduced for the highest energy. For the lowest energy, the fully statistical model has too large branching ratios to states lying below 10 MeV. 

\begin{figure}[h]
\resizebox{\columnwidth}{!}{\includegraphics{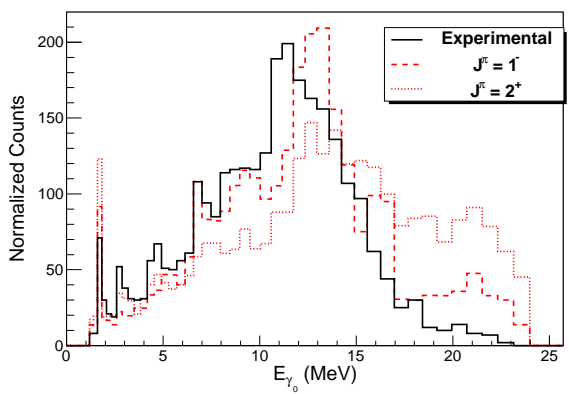} }
\caption{Highest energy $\gamma$-ray spectrum for $E_{c.m.}$= 6.6 MeV (full line) compared to decay scenarii with two different spins in the entrance-channel: $1^-$ (red dashed line) and $2^+$ (red dotted line). Simulations spectra are normalized to the data integral.}
\label{spin_unique_e66}
\end{figure}

As the studied energies correspond to resonant energies, we used another scenario which consists of a unique spin in the entrance-channel. As the spin distributions given in Fig.~\ref{spin_distri} show that entrance-channel spins greater than $5\hbar$ have very small contributions to the fusion cross-sections, we limit the discussion here to entrance spins less or equal to $4\hbar$. Fig.~\ref{spin_unique_e66} displays the simulation results for resonances $J^{\pi} = 1^-$ and $2^+$ and their comparison with the lowest energy data. As for the first scenario, we see that we overall reproduce the shape of the spectrum with the negative parity entrance spins, even if the bump seems to be shifted. Again this is due to too large branching ratios to states lying below the fed states in the experiment and has to be interpreted in terms of structural effects. In the case of a positive parity resonance the large $\gamma$-flux going directly to the g.s. and the $2^+_1$ depletes the region around $E_{\gamma_{0}}=12$ MeV and gives rise to a completely different $\gamma$-spectrum. 

\begin{figure}[h]
\resizebox{\columnwidth}{!}{\includegraphics{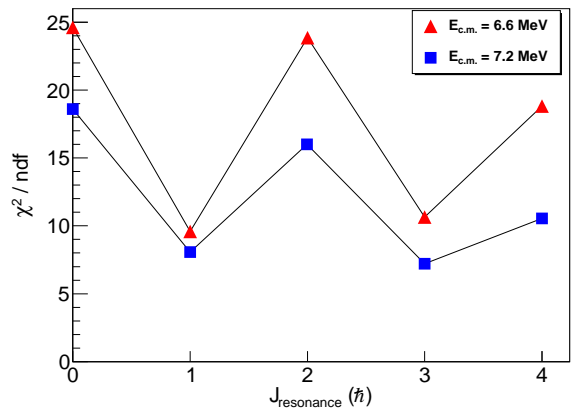} }
\caption{$\chi^{2}$ variation as a function of the spin of the entrance-channel, red triangles correspond to $E_{c.m.}$ = 6.6 MeV and blue squares to $E_{c.m.}$ = 7.2 MeV.  }
\label{chi2_var}
\end{figure}

In order to have a objective criterion to compare simulated and experimental spectra we used  the so-called $\chi^2$ test of homogeneity. Evolution of the $\chi^2$ versus the resonance spin for the two studied energies is given in Fig.~\ref{chi2_var}. Looking at the two curves on Fig.~\ref{chi2_var}, we observe a kind of oscillation between even and odd spins. This can be understood by the fact that for the even spins, corresponding to positive parity in the entrance-channel, decays to the low-lying states of positive parity are mainly achieved by E2 transitions which are not slowed down by any selection rules in this nucleus.

\begin{figure}[h]
\resizebox{\columnwidth}{!}{\includegraphics{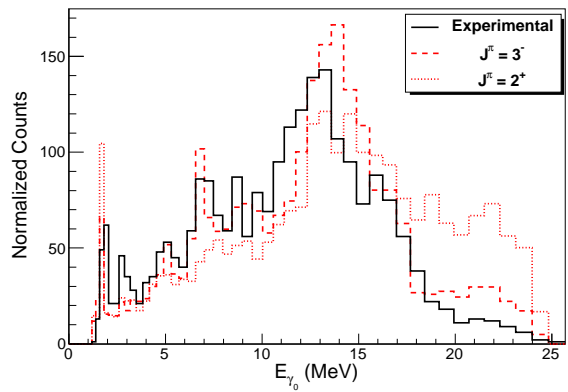} }
\caption{Highest energy $\gamma$-ray spectrum for $E_{c.m.}$= 7.2 MeV (full line) compared to decay scenarii with two different spins in the entrance-channel: $3^-$ (red dashed line) and $2^+$ (red dotted line). Simulations spectra are normalized to the data integral.}
\label{spin_unique_e72}
\end{figure}

For the highest energy for which results are given in Fig.~\ref{spin_unique_e72} the same discussion can be held. The difference is that a $J^{\pi} = 3^-$ resonance in the entrance-channel reproduces the data better than a $J^{\pi} = 1^-$. But as for the lowest energy, data are better reproduced with entrance-channel odd spins. Branching ratios to negative parity states from an negative entrance spin are expected to be favored against branching ratios to positive parity states in resonance $\gamma$-decay. This is partially reproduced by the semi-statistical scenario. We already stressed the importance of particular negative states such as the $3^-_1$ which is the first negative parity state, at higher energies. As the phase space is reduced due to the reduction of the bombarding energy, this particular decay may be more important in the two explored energies than in our previous experimental campaign.

\subsection{Radiative capture cross-section}
The previously unobserved large feeding of intermediate states around 11 MeV allows us to reevaluate the radiative capture (RC) cross-section measured in Ref.~\cite{coll}.
\begin{equation}
\sigma_{RC}=\frac{N_{r}}{N_{t} \cdot N_{i} \cdot T_{a}} \cdot \frac{1}{\epsilon_{det}}\cdot 10^{24} \quad , 
\label{cross_section}
\end{equation}
in which the cross-section, $\sigma_{RC}$, is given in barn. $N_{r}$, $N_{t}$ and $N_{i}$ correspond, respectively, to the number of $^{28}$Si recoils observed at DRAGON focal plane ($\pm 10\%$), the number of $^{12}$C atoms/cm$^2$ ($\pm 10\%$) in the target and the number of incident $^{16}$O per second ($\pm 10\%$). $T_{a}$ denotes the time with beam on target reduced by the data acquisitions dead time (8 and 17 \% for the lowest and highest explored energies, respectively) and is known with an error of $\pm 5\%$. $\epsilon_{det}$ stands for the detection efficiency which takes into account the DSSSD detection efficiency (96.2 $\pm$ 0.1\%) \cite{dragon_com}, the $^{28}$Si$^{8+}$ charge state fraction which corresponds at both energies to 35\% and finally the acceptance and the transport efficiency of DRAGON. This third parameter can be extracted using the GEANT simulations with the best agreement with data and corresponds to 32\% (resp. 36\%) for $E_{c.m.}$ = 6.6 MeV (resp. 7.2 MeV). The error on the acceptance and the transport through DRAGON is of the order of 15\%. 

\begin{figure}[h]
\resizebox{\columnwidth}{!}{\includegraphics{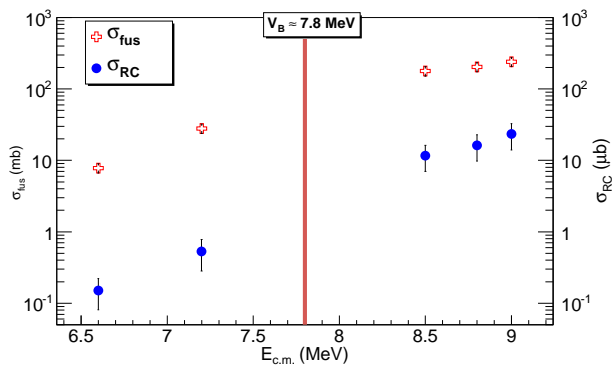} }
\caption{RC cross-sections (blue filled circles) along with fusion cross-sections (red crosses) \cite{cujec}. Error bars for the fusion cross section correspond to the size of the red crosses. RC cross-sections above $V_{B}$ are from a previous study \cite{doro}.}
\label{plot_xsection}
\end{figure}

The obtained cross-sections are  given in Fig.~\ref{plot_xsection} along with data from our previous study \cite{doro} above $V_B$ ($E_{c.m}$ = 8.5, 8.8 and 9.0 MeV) and fusion cross-section at these energies extracted from \cite{cujec}. For the two recently explored energies RC cross-sections are lower than 1 $\mu b$. Furthermore excitation function slopes show that RC cross-section tends to decrease faster than the fusion cross-section below $V_B$. Indeed at $E_{c.m.}$= 9.0 MeV RC cross-section represents $\sim 10 \times 10^{-5}$ of the fusion cross-section, this ratio falls to $\sim 2 \times 10^{-5}$ for the two energies discussed in this paper.

\section{Conclusion}
We have performed a heavy ion radiative capture reaction between two light heavy ions, $^{12}$C and $^{16}$O, leading to $^{28}$Si at two resonant energies below $V_B$. Obtained $\gamma$-spectra display a previously unobserved strong feeding of intermediate states around 11 MeV at these energies. This new decay branch has to be compared to the feeding of the same region observed at higher energies, in order to obtain the evolution of the direct feeding of the different $^{28}$Si known states from the entrance-channel. As the new decay branch can not be fully reproduced by statistical nor semi-statistical decay scenarii this may be the signature of structural effects. With the upcoming new scintillator generation, such as the PARIS project \cite{paris}, a sufficient resolution will be achieved in order to disentangle the feeding of intermediate states around 11 MeV contained in the large bump. New Monte-Carlo simulations with other conditions such as calculated branching ratios with a cluster model for the entrance-channel have to be tested to see if a better agreement with the data can be achieved.

 GEANT simulations have been used to extract the DRA\-GON spectrometer acceptance which is crucial to obtain a value for the radiative capture cross-sections. If new decay scenarii can reach better agreement with data, error bars on capture cross-section will be reduced. This will lead to a better understanding of the fusion reaction below $V_B$ and, more particularly, for astrophysical energies where new resonances have been observed mainly in the neighbouring $^{12}$C+$^{12}$C system in the vicinity of the Gamow window \cite{spil}. 



\end{document}